\documentclass[]{aa}
\usepackage{txfonts}
\usepackage{graphicx}
\usepackage{natbib}
\usepackage{epsfig}
\input psfig.sty

\begin{document}
\title{Non-thermal recombination - a neglected source of flare hard X-rays and fast electron diagnostic
 }
\author{John C. Brown\inst{}\and Procheta C.V. Mallik \inst{}}

\institute {Department of Physics and Astronomy, University of
Glasgow, Glasgow G12 8QQ, U.K.}

\offprints{J.C. Brown, \email{john@astro.gla.ac.uk}}

\authorrunning{Brown and Mallik}
\titlerunning{Fast electron recombination HXRs
}
\date{\today}

 \abstract
 {Flare Hard X-Rays (HXRs) from non-thermal electrons are commonly treated
 as solely bremsstrahlung (free-free = f-f), recombination (free-bound = f-b) being neglected. This assumption is shown to be
 substantially in error, especially in hot sources, mainly due to recombination onto Fe
 ions.}
 {We analyse the effects on HXR spectra $J(\epsilon)$ and electron diagnostics by including non-thermal recombination onto heavy
  elements in our model.}
  {Using Kramers hydrogenic cross sections with effective $Z=Z_{eff}$,
  we calculate f-f and f-b spectra for power-law electron spectra within both
  thin and thick target limits and for Maxwellians with summation over
 all important ions.}
  {We find that non-thermal electron recombination, especially onto Fe,
  must, in general,  be included with f-f for reliable spectral
  interpretation,
  when the HXR source is hot, such as occulted loops containing high ions of Fe (f-b
 cross-section $\propto Z^4$). The f-b contribution is greatest when the electron
 spectral index $\delta$ is large and any low energy cut-off $E_c$ is
 small, because the electron flux spectrum $F(E)$ emitting f-b photon energy
 $\epsilon$ is $\propto (E=\epsilon-V_Z)^{-\delta}$ ($V_Z$ is the ionisation potential) and not  $\propto (E=\epsilon)^{-\delta+1}$ as
 for f-f. The f-b spectra recombination edges mean a cut-off  $E_c$ in
 $F(E)$ appears as an HXR feature at  $\epsilon = E_c + V_Z$, offering
 an $E_c$ diagnostic. For thick target sources, the presence of $E_c$
 appears as edges in $J'(\epsilon)$, not in $J(\epsilon)$,  but it is still detectable.
 Including f-b lowers the $F(E)$ needed for
 prescribed HXR fluxes greatly in some cases;  and even when  small, it seriously distorts $F(E)$ as inferred by
 inversion or forward
  fitting of $J(\epsilon)$ based on f-f alone.}
  {The f-b recombination from
  non-thermal electrons can  be an important
  contributor to HXR spectra, so it should be included in spectral analyses,
  especially for hot sources.
  Accurate results will  require use of better cross sections than ours and
  consideration of
  source ionisation structure.}


\maketitle

\section{Introduction} \label{sect:intro}

    Ever since their first detection (Arnoldy, Kane and
Winckler 1968, Kane and Andersen 1970), flare hard X-ray (HXR) bursts
(photon energies $\epsilon > 10$ keV or so) have been recognised as
an important diagnostic of electron acceleration and propagation
(e.g. Brown 1971, Lin and Schwartz 1987, Johns and Lin 1992). The
large electron flux and power
   imply they play a
substantial role in flare energy budgets and pose challenges for
electron acceleration mechanisms (see recent reviews by, e.g. Vilmer
{\it et al.} 2003, Brown 2005, MacKinnon 2006).  Recent copious high resolution HXR spectral data from the RHESSI mission (Lin
et al 2002) have created the possibility of detailed reconstruction
of source electron spectra (following Brown 1971) offering important
constraints on the electron energy budget and acceleration processes
(Piana {\it et al.} 2003, Conway {\it et al.} 2003, Massone {\it et
al.} 2004, Kontar {\it et al.} 2004, 2005, Brown {\it et al.} 2006).

    In inferring electron flux spectra $F(E$), the HXR radiation
mechanism has always been taken to be f-f collisional bremsstrahlung
of fast electron impacts with atoms and ions, gyrosynchrotron and
inverse Compton radiation being negligible at these energies for
solar magnetic and radiation fields (Korchak 1971). Though included
for thermal electrons in hot (a few keV) plasma, f-b recombination
radiation from non-thermals seems to have been assumed negligible
other than in a preliminary study by Landini, Monsignori and
Pallavicini (1973).  In view of the importance of details in the
photon spectrum $J(\epsilon)$ (photons sec$^{-1}$kev$^{-1}$) for
accurate reconstruction of $F(E$), we re-examine this assumption,
and conclude (cf Mallik and Brown 2007) that it is not valid under
some conditions, which quite commonly exist in some flare HXR source
regions.

It is not the intention of this paper to analyse precisely the
theoretical recombination radiation spectrum from fast electrons
under conditions (e.g. ionisation structure) for specific flares
which are typically both inhomogeneous and time dependent. Rather we
give approximate theoretical estimates of how important it may be
relative to bremsstrahlung under various limiting conditions.
Specifically, we compare the two in the simplest, Kramers,
cross-section approximations, for limiting cases of plasma
ionisation. The recombination emission rate per electron is very
sensitive to the ionic charge, being $\propto Z^4A_Z$ (Kramers 1923)
per plasma proton for hydrogenic ions of charge $Ze$ and number
abundance $A_Z$. Thus the emitted f-b flux and spectrum depend
strongly on the ionisation state, hence the temperature,  of the
plasma where the fast electrons recombine. In practice this will
involve several ionisation stages of several target plasma species
(since $Z^4A_Z$ may be large even for small abundance $A_Z$), which
will vary along the paths of the electrons and be time dependent.

    The paper is organised as follows.  In Section 2 we briefly discuss
relevant processes and the cross-section approximations we use, and
obtain expressions for the total continuum photon spectral
contributions  $j(\epsilon)$ expected from an electron flux spectrum
$F(E)$ from f-f and from f-b processes.  In Section 3, we compare
these for a power-law $F(E)$ with low cut off at $E<E_c$ and for a
shifted power-law, and discuss implications for flare electron
spectra and energy budgets under several limiting plasma ionisation
assumptions. In Section 4 we look at thermal and non-thermal components to show how the relative
importance of each contribution depends on conditions in the flare
by varying parameters around those for a specific real event.
Section 5 discusses the effect of including the f-b contribution on
inverse problem inference of $F(E)$ from $j(\epsilon)$ while Section
6 summarises our conclusions and suggests directions for future
work. Details of some of the equations are given in Appendix A. In Appendix B we discuss the total emission spectra from
extended volumes for thin target, collisional thick target and
thermal cases.

\section{Free-free and free-bound emissivity spectra}

\subsection{General considerations}

In this section, we discuss only local emissivities $j(\epsilon)$
(photons  cm$^{-3}$ sec$ ^{-1}$ per unit $\epsilon$ ). Relativistic
and directivity effects are disregarded ($E,\epsilon \ll m_ec^2$)
since the f-b/f-f ratio is largest at low $E$. Then, if target
atom/ion type $t$ has density $n_t$ and the fast electron flux
spectrum is $F(E)$ (electrons sec$^{-1}$ cm$^{-2}$ per unit $E$),
$j(\epsilon)$ for a collisional radiation process is

\begin{equation}
j(\epsilon)=\Sigma_t j_t(\epsilon)=\Sigma_t n_t
\int_{E_{tmin(\epsilon)}}^\infty
F(E)\frac{dQ_t}{d\epsilon}(\epsilon,E)dE,
\end{equation}
where  $dQ_t/d\epsilon(\epsilon,E)$ is the relevant cross-section
per unit $\epsilon$ for target species $t$ and the integral is over
the range of electron energies relevant to species $t$.

\subsection{Bremsstrahlung}

In the case of f-f (bremsstrahlung), $dQ_t/d\epsilon(\epsilon,E)$ is
essentially the same for any state of ionisation of an atomic
species $Z$ (Koch and Motz 1959), and the $t$ summation in (1) need only
be carried out over elements $Z$ to give, for element abundances
$A_Z$ (by number relative to hydrogen), and total proton (p+H)
density $n_p$,

\begin{equation}
j_B(\epsilon)=n_p\Sigma_Z A_Z
 \int_\epsilon^\infty F(E)\frac{dQ_{BZ}}{d\epsilon}(\epsilon,E)dE,
\end{equation}
where $dQ_{BZ}/d\epsilon(\epsilon,E)$ is the bremsstrahlung
cross-section for element $Z$ and $E_{min}=\epsilon$ since any
free-free transition can only yield a maximum  $\epsilon=E$.  The
bremsstrahlung cross-section per nucleus $Z$ scales as  $Z^2$ and
can be written

\begin{equation}
\frac{dQ_{BZ}}{d\epsilon}=\frac{8\alpha r_e^2
Z^2}{3}\frac{m_ec^2}{\epsilon E}q(\epsilon,E)~, ~\epsilon\le E
\end{equation}
(and zero for $\epsilon >E$). Here $\alpha = e^2/\hbar c$ is the fine
structure constant and $r_e=e^2/m_ec^2$ the classical electron
radius, while $q(\epsilon, E)$ is the ratio of the actual cross
section to the Kramers cross section (Kramers 1923), which is the
factor in front of $q$. While this is only a first approximation,
not suitable for accurate absolute spectral inversion/reconstruction
algorithms (Brown 2005), it will be adequate for the present purpose
of comparing f-f with f-b emission, which we also treat in the
Kramer's approximation. Then (2) and (3) give, for bremsstrahlung,

\begin{equation}
j_B(\epsilon) =\frac{8\alpha
r_e^2}{3}\frac{m_ec^2}{\epsilon}\zeta_Bn_p\int_\epsilon^\infty
\frac{F(E)}{E}dE,
\end{equation}
where

\begin{equation}
\zeta_B=\Sigma_Z \zeta_{BZ}=\Sigma_Z A_ZZ^2
\end{equation}
is the heavy element correction for bremsstrahlung, with
$\zeta_B\approx 1.6$ for
the solar coronal abundances we use - see later.

\subsection{Recombination Radiation}

\begin{table*}
\centering
\scriptsize
\caption{Elements with
their coronal abundances and ionisation potentials at $T \gg 10^8$ K}
\vskip 0.2cm
\begin{tabular}{cccccc}
\hline\smallskip
Element & Z & $A_z$ & $A_zZ^2$ & $A_zZ^4$ & $V_z = Z^2\chi$ (keV)\\
\hline
H & 1 & 1 & 1 & 1 & 0.0136\\
He & 2 & 0.096 & 0.384 & 1.536 & 0.0544\\
C & 6 & 3.57 x $10^{-4}$ & 0.013 & 0.463 & 0.490\\
O & 8 & 8.57 x $10^{-4}$ & 0.055 & 3.511 & 0.870\\
Ne & 10 & 1.07 x $10^{-4}$ & 0.011 & 1.071 & 1.360\\
Mg & 12 & 1.33 x $10^{-4}$ & 0.019 & 2.755 & 1.958\\
Si & 14 & 1.27 x $10^{-4}$ & 0.025 & 4.871 & 2.666\\
S & 16 & 1.61 x $10^{-5}$ & 0.0041 & 1.053 & 3.482\\
Ca & 20 & 8.50 x $10^{-6}$ & 0.0034 & 1.360 & 5.440\\
Fe & 26 & 8.61 x $10^{-5}$ & 0.058 & 39.336 & 9.914\\
Ni & 28 & 6.95 x $10^{-6}$ & 0.0054 & 4.27 & 10.662\\
\hline
 &  &  & $\Sigma = 1.58$ & $\Sigma = 61.2$ & \\
\hline
\end{tabular}
\end{table*}

\begin{table*}
\centering
\scriptsize
\caption{Ionic species of iron at 20 MK}
\vskip 0.2cm
\begin{tabular}{ccccccc}
\hline\smallskip
Element & $Z-z$ & $Z_{eff}$ & $\Phi_{Z_{eff}}$ & $A_z$ & $A_zZ_{eff}^4$ & $V_z = Z_{eff}^2\chi$ (keV)\\
\hline
Fe XXII & 21 & 21.98 & 0.05 & 0.43 x $10^{-5}$ & 1.004 & 6.57\\
Fe XXIII & 22 & 22.61 & 0.14 & 1.21 x $10^{-5}$ & 3.152 & 6.95\\
Fe XXIV & 23 & 23.20 & 0.25 & 2.15 x $10^{-5}$ & 6.232 & 7.32\\
Fe XXV & 24 & 23.77 & 0.56 & 4.82 x $10^{-5}$ & 15.381 & 7.68\\
\hline
\end{tabular}
\end{table*}

    The situation here is more complicated.  Firstly, 2-body radiative
recombination (we neglect 3-body recombination) of a free electron
of energy $E$ to a bound level $m$ of energy $-V(Z,i,m)$ in ionic
stage $i$ yields a photon energy $\epsilon$, which, apart from
quantum uncertainty, is unique, namely:

\begin{equation}
\epsilon=E +V(Z,i,m).
\end{equation}

That is, when a fast electron does recombine, all of its kinetic
energy $E$ {\it plus} $V$ goes into a photon of that energy, in
contrast to bremsstrahlung where photons of all energies
$\epsilon\le E$ are emitted.

Furthermore, for each element $Z$, there is a range of $Z+1$ distinct
ion stages  $i$ each with its own distinct set of energy levels ($m$)
and a set of $Z,i,m$-dependent recombination cross-sections. Thus
recombination collisions of a mono-energetic beam with a
multi-species plasma gives rise to a set of delta-function-like
spectral features at all energies (6) corresponding to elements $Z$,
ionic stages $i$ and levels $m$ . For a continuous electron
spectrum, this yields a continuum photon spectrum that is a sum of
an infinite series of energy-shifted electron flux contributions.
In contrast to bremsstrahlung it does not involve an integral over a
continuum of electron energies.

For a general plasma the basic particle type $"t"$ onto which
recombination occurs is level $m$ of ion stage $i$ of element $Z$
with recombination cross-section differential in $\epsilon$ for that
$t$:

\begin{equation}
\frac{dQ_{Rt}}{d\epsilon}(\epsilon)=Q_{Rt}\delta(E-\epsilon+V_t),
\end{equation}
where $Q_{Rt}$ is the total radiative recombination cross-section
for species $t$ and $\delta(E')$ is the delta-function in energy such
that $\int_{-\infty}^\infty \delta(E')dE'=1$. Then the total
recombination emission spectrum for electron flux spectrum $F(E)$ is

\begin{eqnarray}
j_R(\epsilon) = & n_p\Sigma_t A_t\int_{E_{min}(\epsilon,t)}^\infty
Q_{Rt}(\epsilon,E)\delta(E-\epsilon +V_t)F(E)dE &   \nonumber
\\ = & \Sigma_t A_t n_p Q_{Rt}(\epsilon,\epsilon-V_t)F(\epsilon-V_t), &
\end{eqnarray}
where $A_t$ is the numerical abundance of species $t$ relative to
$n_p$. The forms for $Q_{Rt}$, for general $t$, are complicated and
have to be calculated numerically, as do the values of $A_t$ when
individual ionisation states are considered. However, in the Kramers
approximation (with unit Gaunt factors) there is an analytic
expression for hydrogenic ions, which we will use to estimate
$dj_R/d\epsilon$ compared with $dj_B/d\epsilon$, namely, for
recombination onto level $m$ of the hydrogenic ion of element $Z$
(Kramers 1923, Andersen {\it et al.} 1992, Hahn 1997)

\begin{equation}
Q_{R}=\frac{32\pi}{3{\sqrt
3}\alpha}r_e^2\frac{Z^4\chi^2}{m^3\epsilon E},
\end{equation}
where $\chi=m_e e^4/2\hbar^2$ is the hydrogen ionisation potential.

For an element in its highest purely hydrogenic ion state the
emissivity spectrum would then be

\begin{equation}
j_{RZ}(\epsilon)= \frac{32\pi}{3{\sqrt
3}\alpha}\frac{r_e^2\chi^2Z^4n_z}{\epsilon}\Sigma_m\frac{1}{m^3}
\frac{F(\epsilon-Z^2\chi /m^2)}{\epsilon-Z^2\chi /m^2}
\end{equation}
with the $m$ summation over $m\ge Z(\chi/\epsilon)^{1/2}$, since
recombination to level $m$ yields only photons of $\epsilon \ge
Z^2\chi/m^2$. If the source were so hot that all atoms were almost fully
ionised the total for all $Z$ would be, in this approximation,

\begin{equation}
j_{R}(\epsilon) =\frac{32\pi}{3{\sqrt
3}\alpha}\frac{r_e^2\chi^2}{\epsilon}n_P\Sigma_Z
Z^4A_Z\Sigma_m\frac{1}{m^3}
\frac{F(\epsilon-Z^2\chi/m^2)}{\epsilon-Z^2\chi/m^2}
\end{equation}
for element abundances $A_Z$, with the same $m$ summation limits.

In reality even super-hot coronal flare temperatures are not high
enough to equal the ultra-hot $T\gg 10^8$ K needed to almost fully ionise
all elements into their hydrogenic states, especially Fe, which is
crucial in having by far the highest value of $A_ZZ^4$ - see Table
1. Consequently, to deal accurately with $j_R$ for real flare data,
we would have to take into account the actual ionisation state of
the flare plasma, which varies with time and location (being
radically different in loop tops from loop footpoints), and actual
forms of $Q_R(Z),V_Z$ for non-hydrogenic ion stages.

For our purpose of making first estimates we make the following
simplifying approximations:

\begin{itemize}
\item We treat {\it all} ions using hydrogenic Equations (9) - (11) but with suitably chosen $Z_{eff}$
so that
\begin{equation}
V_Z= Z_{eff}^2 \chi ~~;~~ Q_{RZ}=\frac{32\pi}{3{\sqrt
3}\alpha}r_e^2\frac{Z_{eff}^4\chi^2}{m^3\epsilon E},
\end{equation}
where $Z_{eff}$ makes allowance for screening and other
non-hydrogenic effects. While this will be a rough estimate for some
ions, such approximations are often quite satisfactory for suitable
$Z_{eff}$ (e.g. Hahn and Krstic 1994, Erdas, Mezzorani and Quarati
1993). Here we adopt $Z_{eff}$ such that hydrogenic Equation (12)
gives the correct value of $Q_{RZ}$ as given by exact calculations
such as those of Arnaud and Raymond (1992) for Fe, which is the most
important ion in our analysis. Typically, for an element of atomic
number $Z$ in an ionic state with $z$ bound electrons left,
$Z_{eff}$ is between $Z-z$ and $Z-z+1$
\item Noting that $Q_R \propto 1/m^3$ we include here only recombination
to $m=1$ (in the sense of the lowest empty level of the ion - hydrogenic with $Z=Z_{eff}$ - not of the atom). Higher $m$
contributions are weaker, being $\propto 1/m^3$ though extending to lower
energies with edges at $Z_{eff}^2\chi/m^2$. These should be included
in quantitative data fitting.

\item We focus on situations where the emitting region is near
isothermal and either quite cool, so that only low $V_Z$ element
recombination matters, or very hot so that high $V_Z$ elements
(mainly Fe) are dominant. The former are typically loop
chromospheric footpoints (thick target) and the latter very hot
coronal loops which are either at the limb with their footpoints
occulted, or are so dense as to be coronal thick targets (Veronig and
Brown 2004).

\end{itemize}

Under these conditions, Equation (11) becomes

\begin{equation}
j_{R}(\epsilon)=\frac{32\pi}{3{\sqrt
3}\alpha}\frac{r_e^2\chi^2}{\epsilon}n_p\Sigma_{Z_{eff}}
Z_{eff}^4A_{Z_{eff}}
\frac{F(\epsilon-Z_{eff}^2\chi)}{\epsilon-Z_{eff}^2\chi},
\end{equation}
where $A_{Z_{eff}}=A_Z\Phi_{Z_{eff}}$ with $\Phi_{Z_{eff}}$ the
fraction of atoms of element $Z$ in ionic state $Z_{eff}$.

 Note that, since there is no integration over $E$ here,
if $F(E)$ contains a sharp feature at an electron energy $E_*$, such
as a low or high $E$ cut-off, this will appear in the recombination
contribution to the photon spectrum $j(\epsilon)$ as a series of
sharp features at photon energies
$\epsilon(m,Z,E_*)=E_*+Z_{eff}^2\chi/m^2~;~m=1,\infty$ for every ion
$Z$ present. The same is true for broad features like smooth bumps or dips. This is in contrast with the bremsstrahlung
contribution, in which such features are smoothed out by integration
over $E$. Thus, even if $j_R\ll j_B$, it may have an important effect
in inferring $F(E)$ from $j(\epsilon)$ since this essentially
involves differentiating $j(\epsilon)$ (Section 5).

\subsection{Element parameters and flare plasma ionisation}

 The heavy element correction for bremsstrahlung, $\zeta_B$,  is almost independent of ionisation state (since
the bremsstrahlung cross sections for atoms and ions of the same $Z$
are essentially the same), being $\zeta_B\approx 1.6$ for solar
abundances. On the other hand $\zeta_{RZ_{eff}}=Z_{eff}^4A_{Z_{eff}}$
depends on the number of empty ion levels available for
recombination. The importance of fast electron recombination
radiation thus depends on the state of ionisation of the plasma in
which the fast electrons  are moving, which is primarily a function
of plasma temperature $T$.

 In Table 1 we show the values of $Z$, $Z^2A_Z=\zeta_{BZ}$,
$Z^4A_Z=\zeta_{RZ}$, $V_Z$ for various elements/ions whose
$\zeta_{RZ}=Z^4A_Z$ is large enough to be significant, if the
element is sufficiently ionised. With $\zeta_{RZ}\approx 40$ for
FeXXVI, Fe is by far the most important if conditions are such that
it is highly ionised. The $kT$ where maximum ionisation of an ion
stage is reached is typically of the order $0.1 Z_{eff}^2\chi$ to
$Z_{eff}^2\chi$. In Table 2 we show more detailed values for several
stages of ionisation of Fe (XXII-XXV, i.e. 21+ to 24+) with the
appropriate $A_{Z_{eff}}=A_{Z_{eff}}\Phi_{Z_{eff}}$ for each of
these Fe ionic states for the typical coronal flare case of
$T=2\times 10^7$ K. These are taken from Arnaud and Raymond (1992)
as are the actual ionisation fractions we adopt later (Section 4)
for the temperatures of the real flare we consider.

The radiative recombination coefficients give $Z_{eff}$, which differ
slightly from the $Z$ values, as mentioned in Section 2.3. For the
2002 April 14 event, to which we return later, the peak flare
temperature was 19.6 MK, $\sim 5\%$ of the iron appearing as Fe XXII
($Fe^{21+}$), $\sim 14\%$ in the Fe XXIII ($Fe^{22+}$) state,
$\sim25\%$ appearing as Fe XXIV and $\sim56\%$ as Fe XXV. The
respective $Z_{eff}$ values are 21.98, 22.61, 23.20 and 23.77.

 Broadly speaking in typical flare/micro-flare conditions we
can consider the following $T$ regimes:

\begin{itemize}
\item At $T\le 10^4$  K ('cold') even H and other low $V_{Z_{eff}}$ ions are neutral so $\zeta_{RZ}\approx 0$ for all $Z$.
This would be typical of  very dense cool chromospheric thick target
footpoints relevant to deeply penetrating electrons.
\item For $10^5 \le T\le 10^6$ K ('cool') the predominant elements ionised are H, O, Mg, Si giving
$\Sigma_Z \zeta_{RZ}\approx 15$. This is most relevant to upper
chromospheric dense warm plasma reached by moderate energy thick
target electrons.
\item At $T\ge 10^7$ K ('hot') Fe is well ionised up to about Fe XXV giving $\Sigma_Z \zeta_{RZ}\approx 50$.
This is relevant to the hot 'coronal' loop regime, hence either to
(i) typical upper (SXR) flare loops of moderate density (thin
target) whose HXR emission is seen in isolation either by HXR
spectroscopic imaging or volume integrated but with the cool
footpoints occulted because they are over the solar limb; or (ii)
cases of coronal thick target loops (Veronig and Brown 2004) where
the upper loop density suffices to stop the fast electrons
collisionally.
\end{itemize}

\section{Local (thin target) HXR spectra of f-f and f-b for power-law $F(E)$ with cut-off}
\subsection{Basic expressions for $j_B,j_R$}

 To estimate how the fast electron recombination
$j_R(\epsilon)$ compares with bremsstrahlung $j_R(\epsilon)$, we first
consider the commonly studied case of a power-law with a low energy
cut-off

\begin{equation}
\label{definePL}
 F(E) = (\delta-1)\frac{F_c}{E_c}\left(\frac{E}{E_c}\right)^{-\delta}~~;~~
 E\ge E_c,
\end{equation}
where $F_c$ is the total electron flux at $E\ge E_c$.
 Then, from
Equations (4) and (14), we obtain for f-f emission

\begin{eqnarray}
\nonumber j_B(\epsilon)  = &
\frac{\delta-1}{\delta}\frac{8\alpha\zeta_B}{3}\frac{m_ec^2r_e^2}{\epsilon}
\frac{n_pF_c}{E_c} &\\
\nonumber & \times ~~\left[\frac{\epsilon}{E_c}\right]^{-\delta}; &
\epsilon \ge E_c \\ & \times ~~1; & \epsilon < E_c,
\end{eqnarray}
while for f-b emission from an ion of effective charge $Z_{eff}$,

\begin{eqnarray}
\nonumber j_{RZ_{eff}}(\epsilon) =& (\delta-1)\frac{32\pi
\zeta_{RZ_{eff}}}{3^{3/2}\alpha}\frac{r_e^2
\chi^2}{\epsilon}\frac{n_pF_c}{E_c^2}  & \\
\nonumber & \times ~~\left[\frac{\epsilon-Z_{eff}^2\chi}{E_c}
\right]^{-\delta-1}; & \epsilon \ge E_c+Z_{eff}^2\chi \\
& \times ~~0; & \epsilon  < E_c+Z_{eff}^2\chi,
\end{eqnarray}
where
\begin{equation}
\zeta_{R_{Z_{eff}}} = A_{Z_{eff}} Z_{eff}^4.
\end{equation}
So the total for all relevant $V_{Z_{eff}}$ is

\begin{equation}
j_{R}(\epsilon) = \Sigma_{Z_{eff}\ge [(\epsilon-E_c)/\chi]^{1/2}}
j_{RZ_{eff}}(\epsilon).
\end{equation}

\subsection{Ratio of $j_R$ to $j_B$}

For this truncated power-law case, the ratio of f-b to f-f
emissivity is

\begin{eqnarray}
\nonumber \Psi = \frac{j_R(\epsilon)}{j_B(\epsilon)}& \frac{2\pi\delta}{\sqrt 3} \frac{\chi}{\epsilon}\Sigma
_{Z_{eff}^2>(\epsilon-E_c)/\chi)}~~
\frac{\zeta_{RZ_{eff}}}{\zeta_B}\left[1-\frac{Z_{eff}^2\chi}{\epsilon}\right]^{-\delta-1}
& \\ & \approx \frac{0.25(\delta/5)}{\epsilon(keV)}\Sigma
_{Z_{eff}^2>(\epsilon-E_c)/\chi)}~~
\frac{\zeta_{RZ_{eff}}}{\zeta_B}\left[1-\frac{Z_{eff}^2\chi}{\epsilon}\right]^{-\delta-1},
&
\end{eqnarray}
where each term in the summation is zero at $\epsilon <
E_c+Z_{eff}^2\chi$.

For $\epsilon \gg E_c,\Psi \rightarrow
0.25\Sigma_{Z_{eff}}A_{Z_{eff}} Z_{eff}^4/\epsilon$(keV). In pure ionised H
($\Sigma_Z\zeta_{RZ}=1)$ this is only $2.5 \%$ at 10 keV.
 This rather small value of $\Psi$
must be the origin of the conventional wisdom that f-b can be
ignored compared to f-f emission at HXR energies. However, this
notion neglects several crucial facts:

\begin{itemize}
\item At high coronal flare temperatures, where all elements are highly ionised,
in plasmas of cosmic chemical abundances, heavy elements are the
main contributors to the $A_ZZ^4$ sum. For  the extreme ultra-hot
case of near-total ionisation of all $Z$, and for modern solar coronal
abundances the $\Sigma_Z$ factor is
$\approx 61.2$, mainly due to Fe as discussed in Section 2.4 - see
Tables 1 and 2. Note that Fe coronal abundance, for example, has been assumed to be $2.9$ times photospheric Fe abundance (Feldman {\it et al.} 1992). Even higher factors of about 4 have been suggested (Dennis, personal communication).
\item At lower $\epsilon$ the contribution from each $Z_{eff}$ rises steeply
to a sharp recombination edge at $\epsilon = E_c+V_Z$, where the flux can be
large, especially if $E_c$ is small and $\delta$ large.
\item At the edge, the [ ] factor in Equation (19) goes to
$[1+Z_{eff}^2\chi/E_c]^{\delta+1}$. This is because the flux of
electrons emitting recombination photons of energy $\epsilon$ is not
the flux of those at $E\ge\epsilon$, as for bremsstrahlung, but of those at
$E=\epsilon-Z_{eff}^2\chi$. Consequently $\Psi$ is not negligible
even at $\epsilon \gg E_c$. For fully ionised Fe alone, this factor is
$\approx [1+10/E_c$(keV)$]^{\delta+1}$, which, for $\delta =5$ and at
$\epsilon = 10$ keV, is 64, 11.4, 5.5 for $E_c=$ 10, 20, 30 keV
respectively. Even for lower stage Fe ions (e.g. XXV), common in
flare coronal loops, evidently recombination must be a significant
contributor to the HXR emission in those parts of the flare.

\end{itemize}

\begin{figure*}
\centering

\includegraphics {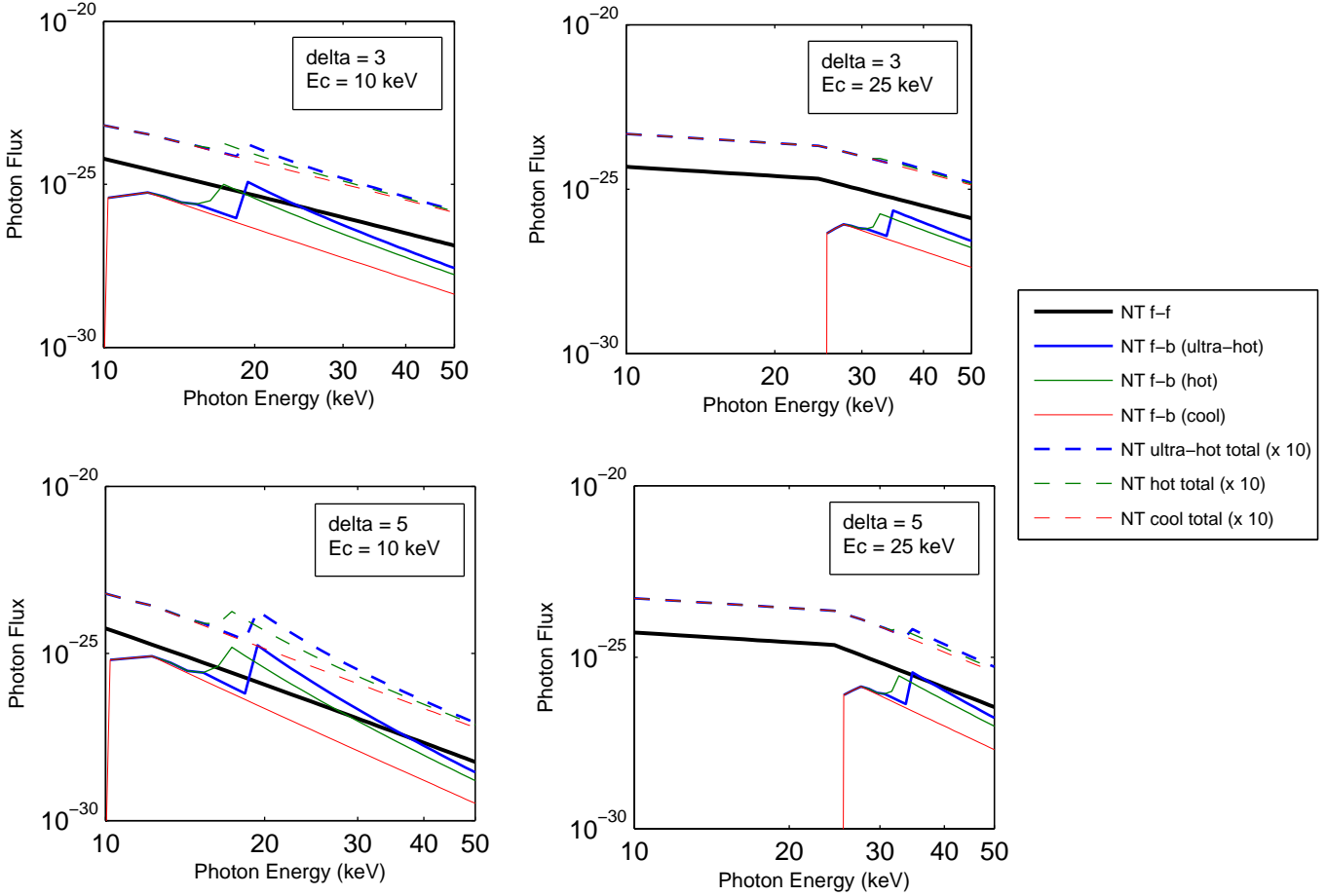}
\caption{Actual shapes of non-thermal f-b and f-f spectra for different temperature regimes and non-thermal electron parameters. Note that the cool, hot and ultra-hot totals are almost identical and the dashed curves nearly indistinguishable for $E_c = 25$ keV.}
\end{figure*}

\begin{figure*}
\centering
\includegraphics {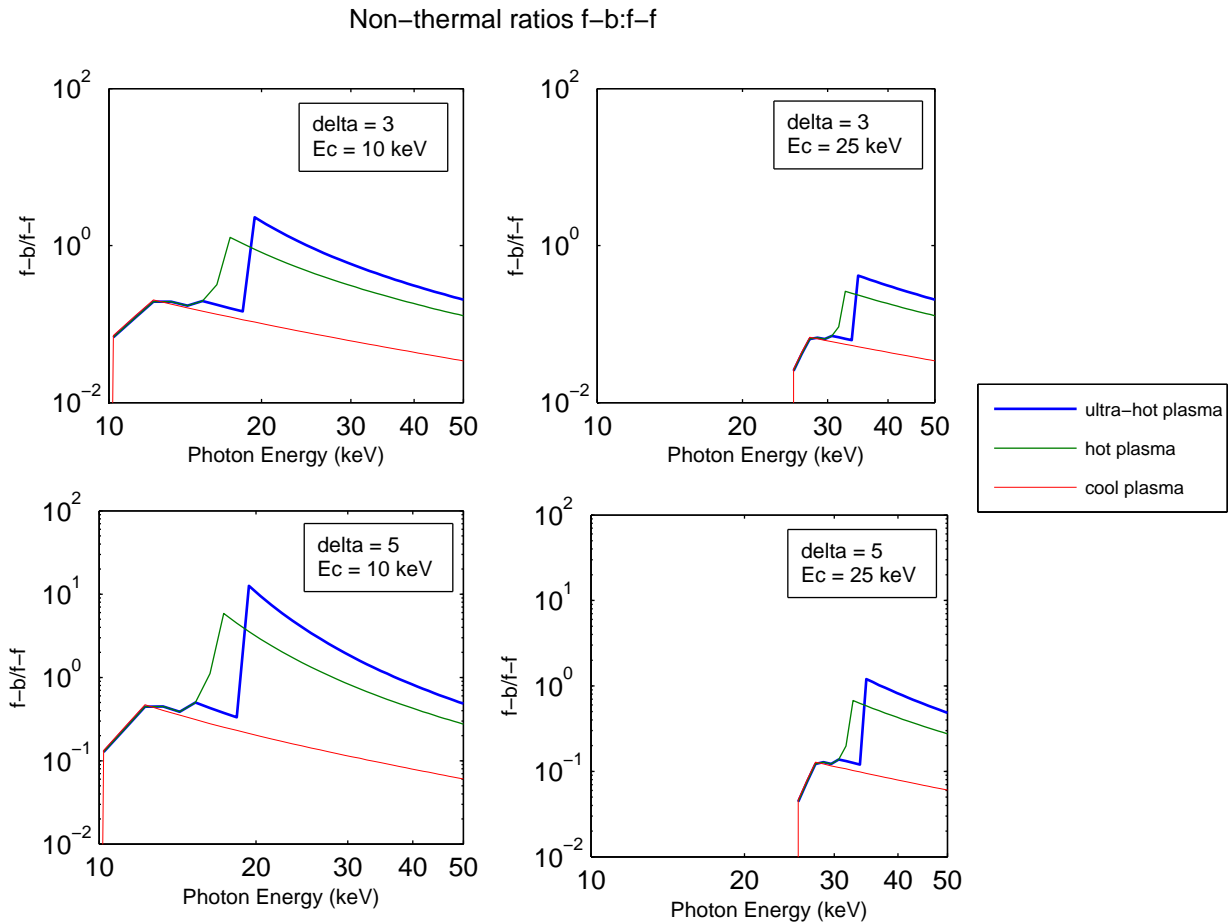}
\caption {Photon flux ratio of non-thermal f-b to f-f emission for different temperature regimes and parameters. Line styles have the same meaning as in Figure 1.}
\end{figure*}

\begin{figure*}
\centering
\includegraphics {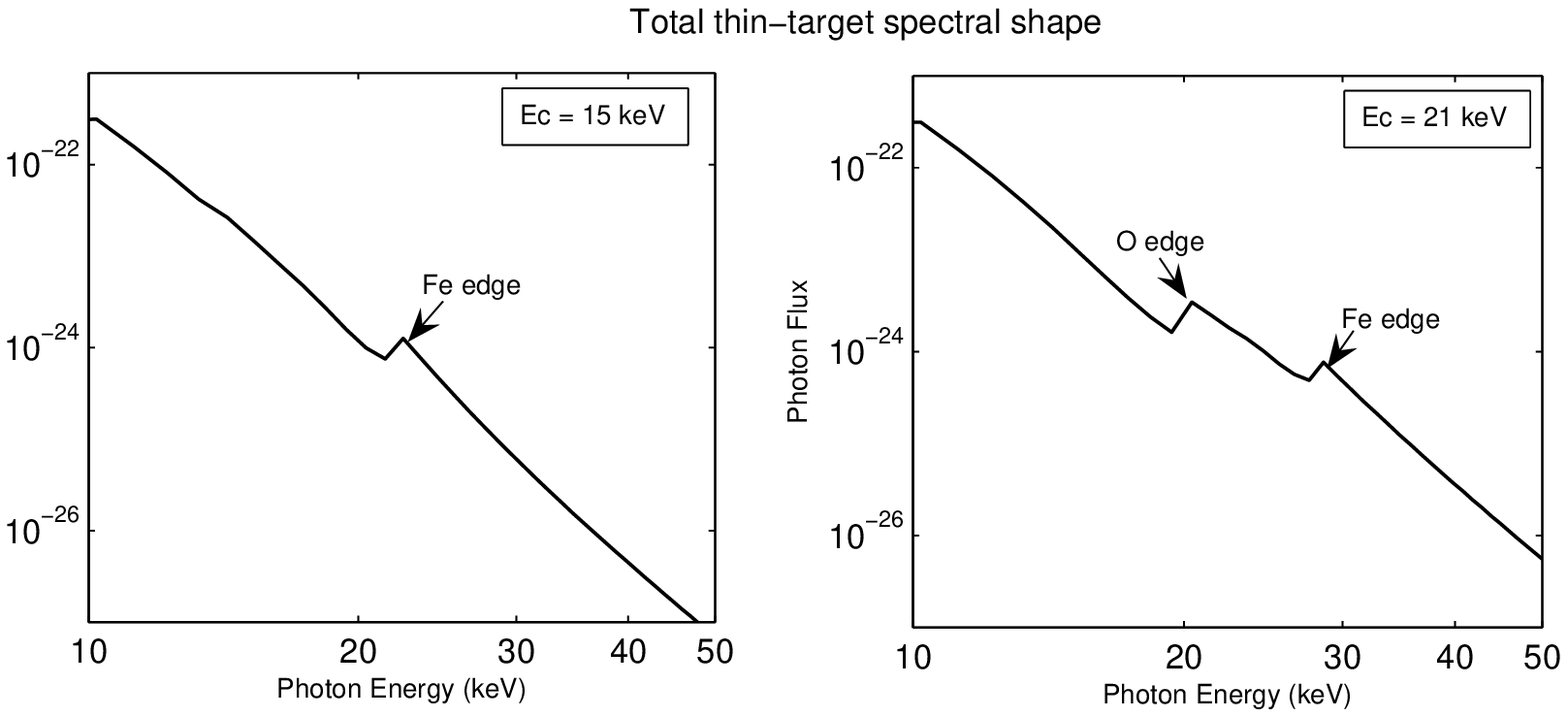}
\caption{Spatially localised spectra from a loop with the 2002 April 14 event plasma parameters for two values of $E_c$. The left plot shows a very
distinct iron edge at $\approx 22$ keV $(=E_c + V_{Fe 24+})$ and a much
less predominant oxygen edge at $\approx 15$ keV $(=E_c)$, whereas the
second plot shows very distinct oxygen  ($\approx 21$ keV
$(=E_c)$) and iron ($\approx 28$ keV) edges. This shows the value of recombination as an $E_c$ diagnostic. The 'edges' appear to be of finite slope because of the finite (1 keV) resolution used.}
\end{figure*}

\begin{figure*}
\centering
\includegraphics {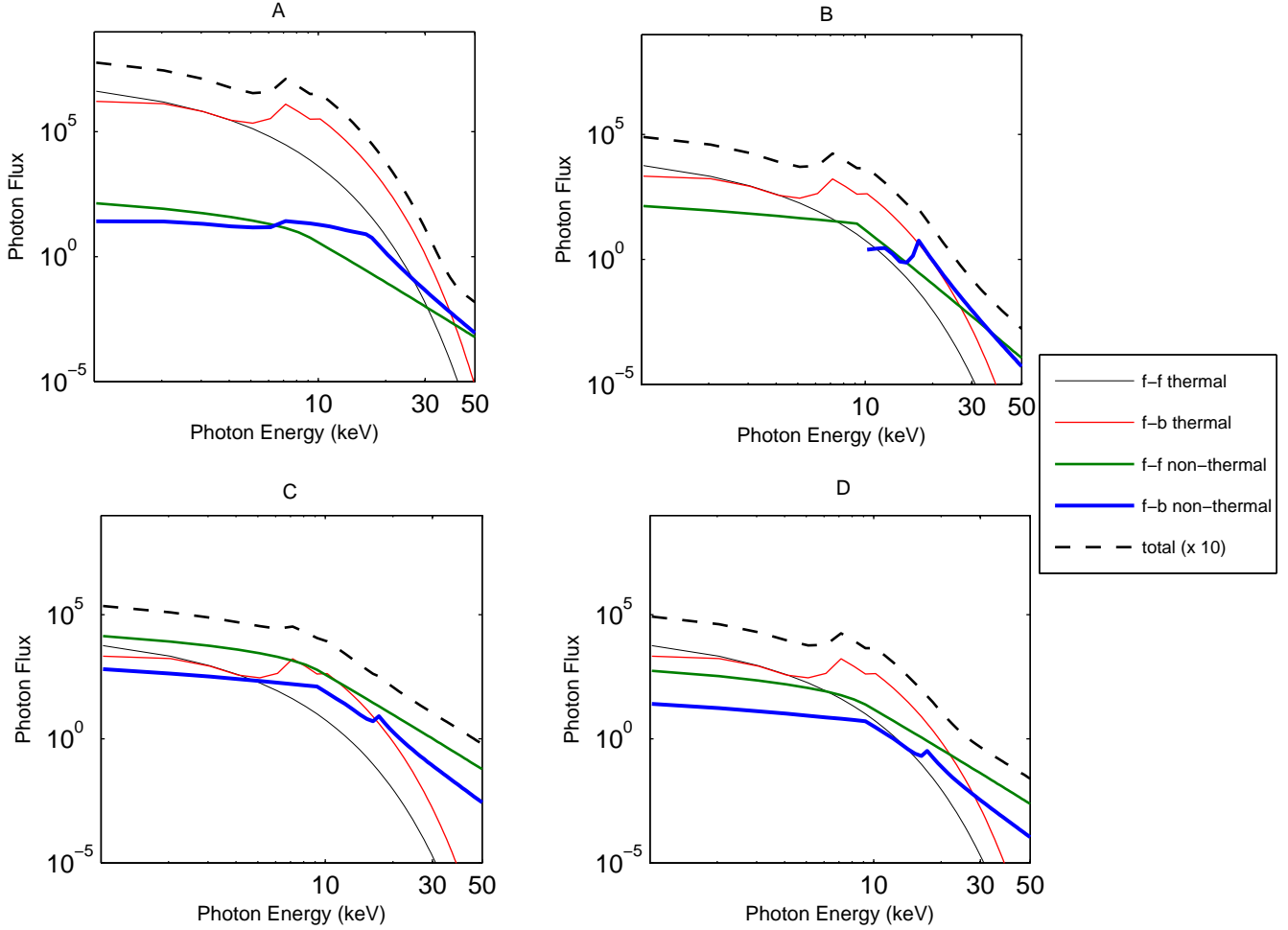}
\caption {The spectral components for 4 different
hypothetical situations. We show these spectra by varying the parameters around the results in
the Veronig and Brown 2004 paper that analyses the coronal thick target 2002
April 14 event. In all cases we keep the same values of $\delta = 6.7$,
$E_c=10$ keV and $T=19.6$ MK.
Plot A is for the thick-target coronal case with the actual
event parameters $n_p,{\cal F}_{oc}$ according to Veronig and Brown.
Plot B was obtained for the same event parameters
but with $n_p$ reduced 25 times to make the loop collisionally thin above
$10$ keV and with footpoint emission occulted. The injection rate is the
same as Plot A so the density fraction of fast electrons is 25 times
higher. The non-thermal emission is down by 25 times while the thermal is down by a factor of 625.
Plot C is the same as B but with cold thick target
footpoints included. The cold footpoint emission (motsly f-f) is dominant.
 Plot D is the same as C, but with an injection rate  reduced by a factor of
 25 so that the density fraction of fast electrons is the same as in Plot A.
 Evidently the detectability of the f-b contribution and of associated
 features in $F(E)$ is sensitive to plasma parameters and observing
 conditions/geometry.}
 \end{figure*}

\begin{figure*}
\centering
\includegraphics {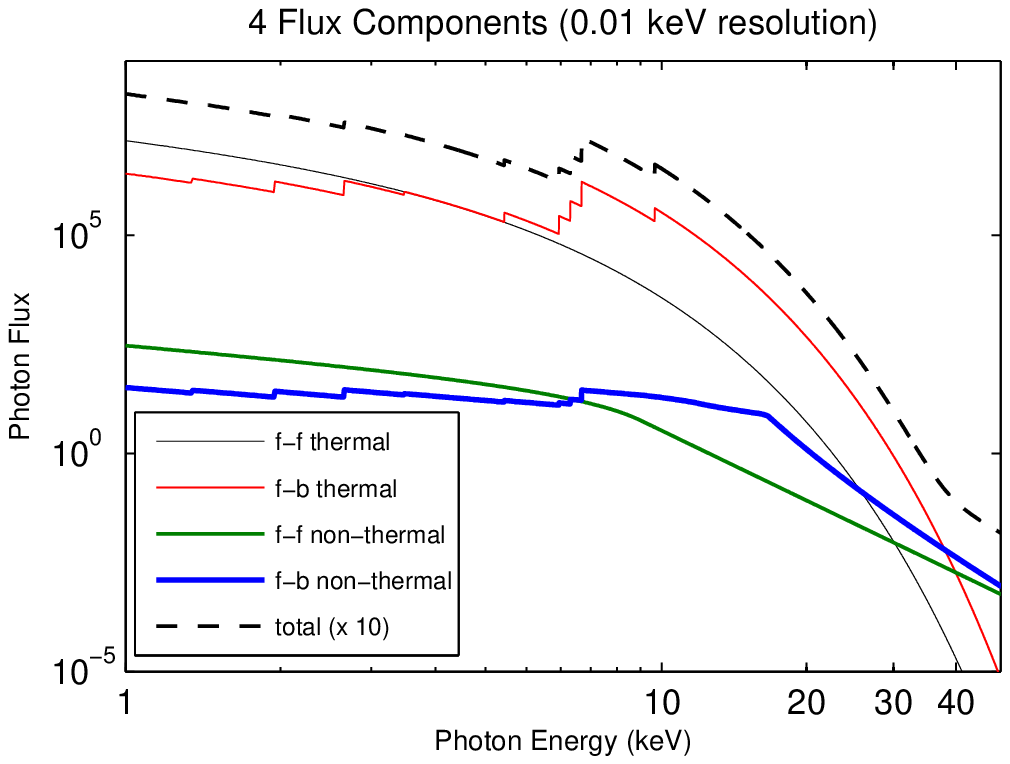}
\caption {Spectral components for a resolution of 0.01 keV. The f-b edges of all elements involved are clearly noticeable. The parameters are $T = 19.6$ MK, $E_c = 10$ keV and $\delta = 6.7$. This plot can be compared to Plot A of Figure 4, which has the same parameters but for 1 keV binning resolution.}
\end{figure*}

\subsection{Typical results in limiting regimes}

N.B. All spectrum figures in this paper (except Figure 5) have been plotted for a bin-width of 1 keV to match RHESSI's spectral resolution. However, in Figure 5 we use 0.01 keV resolution so as to compare it with Plot A of Figure 4 to see how f-b edges would look if they were observed at a higher resolution. The 1 keV binning smears out a lot of the edges of different elements that are clearly noticeable in Figure 5. Hence in Figures 3 and 4, the edges are not 'infinitely' steep as they should be; this is evident in Figure 5 where they do look 'infinitely' steep due to the finer resolution. Also important to note is that the features seen in Figures 3, 4 and 5 are recombination edges and not spectral lines. None of the figures in this paper includes spectral lines - leaving them out shows more clearly where f-b edges exist in the HXR continuum.

In Figure 1 we show for $\delta=3, 5$ the actual spectral shapes
for  $E_c=10, 25$ keV respectively in plasmas of normal solar coronal abundances, which are: ultra-hot
($T \gg 10^8$ K; Fe is nearly fully ionised), hot ($T = 2\times 10^7$ K; Fe well ionised up to Fe XXV) and cool ($T = 10^6$ K; elements up to Si are almost fully ionised).
In Figure 2 we show the ratios $\Psi(\epsilon)$
for the ultra-hot, hot and 'cool' cases, respectively.
The following key features of the hot thin target situation are
apparent from these Figures:

\begin{itemize}
\item The peak non-thermal f-b contribution, in each hot or ultra-hot case shown,
adds at least 50\% to the usual f-f one and in some cases
($\delta=5,E_c=10$ keV) is up to 10 times greater (1000\% increase)
even when only ions up to Fe XXV are present. This is essentially
due to the high abundance of Fe - much higher than thought when
recombination spectra were first discussed (Culhane 1969, Culhane
and Acton 1970).  In appendix A we evaluate the efficiency with which f-b yields HXRs compared to f-f, and also derive the ratio $\Psi$ for the case of a smooth $F(E)$ with no cut-off. This proves, that in a hot enough plasma, far less electrons and power are needed than is found when only f-f is included and that, for smooth $F(E)$, $\Psi$ is largest for large $\delta$ and low $E$ spectral roll-over.
\item In the 'cool' case ($T\approx 10^6$ K) of elements
up to Si almost fully ionised, the f-b contribution is smaller but not in
general negligible. For example, in the bottom left panel of Figure 1
($\delta=5, E_c=10$ keV), f-b is about 30 \% of f-f at 15 keV
energies. This is amply large enough to have a major impact on
inferring $F(E)$ by inversion or by forward fitting (Section 5).
\item In hot plasma, Fe is by far the most important contributor of
recombination radiation.
\item The peak ratio of f-b to f-f increases as $\delta$ is increased
and/or $E_c$ is decreased. This is because f-b photons of energy
$\epsilon$ are emitted by electrons of energy $E-V$ which have flux
$F(E-V)\propto (E-V)^{-\delta}$ which is greatest when the minimum
$E=E_c$ is smallest, $V$ is largest and the steepness $\delta$
greatest.
\item Recombination edges are apparent for the elements with the
highest values of $A_{Z_{eff}}Z_{eff}^4$ - Fe, Si, Mg and O and at
energies $\epsilon = Ec+Z_{eff}^2\chi$, thereby creating the
possibility of finding the location of a low energy cut-off $E_c$
should one exist.
\item The harder asymptotic $\gamma=\delta+1$ for f-f compared with
$\gamma=\delta+2$ for f-b (Equations (15) and (16)) results in an
upward 'knee' in the total spectrum clearly visible in Figure 1 for
$E_c=10$ keV but also present for higher $E_c$ outside the
$\epsilon$ range of the Figure. This could be an important signature
in data of a substantial f-b contribution.

\end{itemize}

 While the edge locations  and the spectral shape
trends will be roughly right, our use of the hydrogenic and
$Z_{eff}$ approximations, and adoption of unit Gaunt factors, mean
that these curves/analytic forms can only be used for approximate
quantitative fitting of real data. As far as we are aware (Kaastra,
personal communication) the Gaunt factors, rates etc. have only ever
been systematically evaluated for Maxwellian $F(E)$ and sometimes
for forms which can be written as sums of these (such as pure
power-laws with no cut-off), and some occasional consideration of
specific non-thermal spectra (e.g. Landini, Monsignori Fossi and
Pallavicini 1973). Comparison of our Maxwellian results, in
the unit Gaunt factor Kramers approximation, with those of
Culhane for the same parameters shows the necessary corrections in
the Maxwellian case to be significant for quantitative comparison
with real data. In addition, in real cases the non-thermal emission
will always be superposed on thermal contributions (especially
important for the very hot plasmas of special interest here) and
also in many cases on a thick target non-thermal contribution
(unless this is from occulted footpoints), from the flare volume as a
whole. In Appendix B we derive the generalisation of the above equations to the various cases involved in real flares, viz. finite volume thin targets, Maxwellian plasmas and thick targets for use in Section 4, where we evaluate the sum of all these contributions for a specific case.

\section{Some practical case study results derived from a real flare}

We saw above and in the appendices that the most favourable conditions for a substantial
recombination contribution are when the maximum possible amount of the
observable HXR source is a hot plasma (e.g. loop) at SXR
temperatures. High density maximises the emission measure but may
make the source/loop collisionally thick and smear recombination
edge spectral signatures of low energy cut offs.  So an optimal case
could be a loop which is just tenuous enough to be collisionally thin
and for which the cool dense thick target footpoints are occulted.
(Footpoint removal by imaging is limited by RHESSI's dynamic range).
Such sources will have a strong HXR source in the coronal loop. One
such event was adopted as a basis for a case study, starting from
the real event parameters. This was the 2002 April 14 event, which
Veronig and Brown (2004) showed to be a hot, dense, collisionally
thick loop with a strong coronal HXR source and no footpoints up to
at least 60 keV. Thus the hot coronal source of non-thermal f-b
emission was not diluted by cold footpoint thick target f-f emission
though the f-b edges were smeared because the hot loop itself slowed
the fast electrons to rest. In Figure 3 we show the theoretical spectrum from a hypothetical resolved part of the coronal loop for two $E_c$ values. We have evaluated the theoretical
thermal, non-thermal and the whole volume hypothetical total $J_B(\epsilon),J_R(\epsilon)$ (from Sections 2-3 and Appendix B) for
such a loop, based on our approximate Kramers expressions, in three
loop parameter regimes (Figure 4):

\begin{itemize}
\item Plot A: With the actual hot thick target loop parameters found by Veronig and Brown, namely
 $\delta=6.7$; $T=19.6$ MK; $L=45 \times 10^8$ cm; $A=19.1 \times 10^{16}$ cm$^2$;
 $n_p=10^{11}$ cm$^{-3}$; $N=4.9\times 10^{20}$ cm$^{-2}$; ${\cal
 F}_1=5\times 10^{35}$ sec$^{-1}$ above $E_1=$ 25 keV. The total $J$ is dominated by thermal f-b and f-f at low $\epsilon$ but thick-target f-b at medium $\epsilon$ and thick-target f-f at high $\epsilon$. Locally within the loop volume,
 if this were spatially resolved, the spectrum $j$ would be like
 those in Figure 3, where edges are clearly visible in positions
 corresponding to cut-off energies of 15 and 21 keV. At a higher resolution, these edges would look similar to the edges shown in Figure 5. Should such edges be found in data, they can diagnose the all-important $E_c$ parameter.

\item Plot B: With the actual parameters found by Veronig and Brown except
with $n_p$ reduced by a factor of 25 so that the loop is collisionally
thin above about 10 keV but with the footpoints hidden (limb
occulted) so there is no cold thick target contribution. In this
case the thermal emission is also much reduced because $EM=2n_p^2AL$
is down by a factor of 625. Somewhere between this and the first
case should be the optimum condition for seeing maximum f-b
contribution.

\item Plot C: The same as B but with the dominant cold footpoint thick target emission added
to show its diluting effect.

\item Plot D: The same as C but with a reduced injection rate and so the thermal is more dominant than in C and this alters the total spectral shape a little bit.

\end{itemize}

The upward 'knee' apparent in Figures 4 A,B at around 40 keV due
to the transition from a f-b to a f-f dominated spectrum (cf.
Section 3 and Figure 3) is rarely seen in data but may be present in
some events (Conway {\it et al.} (2003)). A statistical survey of a
large sample of events should shed light on conditions where
non-thermal f-b is important. Also note that an upward 'knee' is present at the transition from a thermal- to a non-thermal-dominated spectrum. The position of this knee depends on the plasma temperature and may interfere with the f-b to f-f 'knee', which depends mainly on the $E_c$ parameter. Hence, although for certain parametric conditions one may be able to notice two separate upward 'knees', if $E_c$ is low and $T$ is high, the 'knees' may occur at similar $\epsilon$ and may not be distinguishable in real data.

\section{The inverse problem - effect of f-f on $F(E)$ inferred from
data on $j(\epsilon)$}

We note again that, since even the thin target $j_B$ involves an
integral over $E$ while $j_R$ does not, any sharp features in $F(E)$
would be smoothed out in the bremsstrahlung contribution to the
photon spectrum but not in the recombination contribution.
Consequently, an important way to study the effect of including f-b on
the required properties of $F(E)$ is to consider it as an inverse
problem (Craig and Brown 1986) to infer $F(E)$ from observed
$j(\epsilon)$. Here we consider the following experiment for the
thin target case. (Thick target and thermal cases always involve even
greater error magnification - Brown and Emslie 1988). Generate the
total $j(\epsilon)$ including f-b as well as f-f from a specified
$F_1(E)$ and evaluate the $F_2(E)$ which would be erroneously
inferred by solving the inverse problem ignoring the  presence of
the f-b term, as is currently done in all HXR data analysis, whether
by inversion or forward fitting.

\begin{figure*}
\centering
\includegraphics {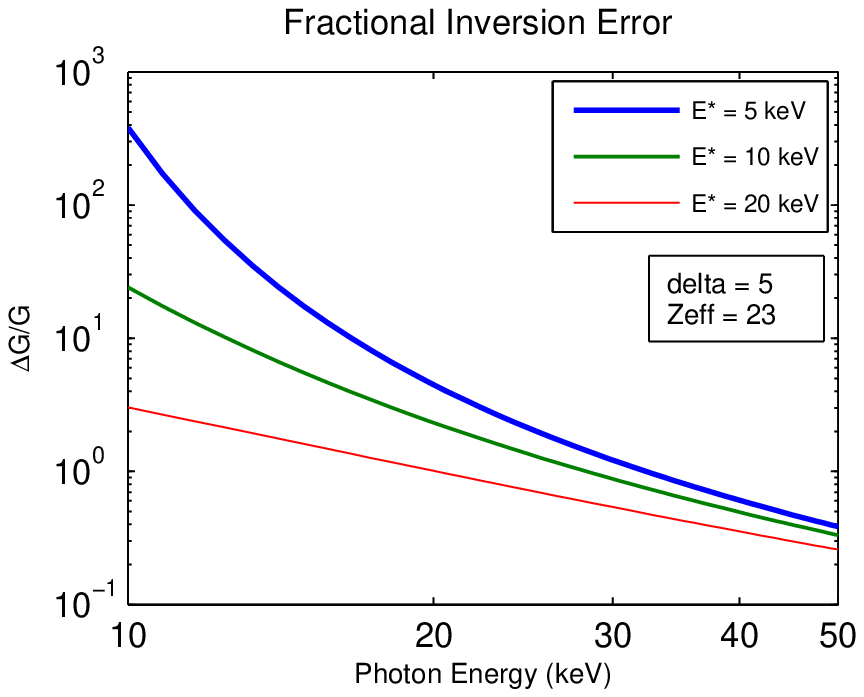}
\caption{Fractional error ($\Delta G/G$) in G (Equation (25)) as
discussed in Section 5 for $E_* = 5, 10, 20$ keV respectively for a shifted power-law due to inference of G from H
ignoring the presence of recombination.}
\end{figure*}

By (4) and (11) the total f-f + f-b emission spectrum $dJ/d\epsilon$
 from a homogeneous volume $V$ can be written

\begin{equation}
H(\epsilon)= \int_\epsilon^\infty G(E)dE + D\Sigma_{Z_{eff}\le{\sqrt
{\epsilon/\chi}}}Z_{eff}^4A_{Z_{eff}}G(\epsilon-V_{Z_{eff}}),
\end{equation}
where
\begin{equation}
H(\epsilon)=\frac{3}{8\alpha
r_e^2}\frac{1}{\zeta_Bm_ec^2n_pV}\epsilon\frac{dJ}{d\epsilon}; \hspace{0.2cm}G(E)= F(E)/E
\end{equation}
and $D$ is as given in Equation (A.2). If we ignore the second
(recombination) term in Equation (20), as has always been done in
the past, for the Kramers f-f term, the inverse is just (Brown and
Emslie 1988)

\begin{equation}
G(\epsilon) = - H'(E).
\end{equation}

The neglect of the second term can be thought of as an 'error'
$\Delta H$ in our data and if we apply inversion formula (22) to this 'data', ignoring the recombination 'error' we get a resulting error
$\Delta G$  in the inferred $G$ given by
\begin{eqnarray}
\Delta G(E) = \frac{F_2(E)-F_1(E)}{E}\\ \nonumber =-D\Sigma_{Z_{eff}\le{\sqrt
{\epsilon/\chi}}}Z_{eff}^4A_{Z_{eff}}G'(E-V_{Z_{eff}}).
\end{eqnarray}

It is at once clear that any sharp change in $j(\epsilon)$ i.e. in
$H(E)$, such as the presence of f-b edges, however small, can have a
very large effect on the inferred $F_2(E)$. (If the inverse problem
is addressed for more realistic smoother forms of f-f cross section
than Kramers, the 'error magnification' is in general even larger -
Brown and Emslie 1988, Piana {\it et al.} 2000). For a power law $F$
with cut off around say 20 keV, analytically speaking this
expression gives infinite negatives in $\Delta G(E)$ at the spectral
edges around 30 keV (for Fe). However when smoothed over a few keV
and added to the f-f term the result would be a 'wiggle' in the
$F(E)$ solution in the 30-40 keV range. This is just where enigmatic
features have been reported in some RHESSI spectra and variously
attributed to the effects of photospheric albedo (Kontar {\it et
al.} 2006), possibly pulse pile up (Piana {\it et al.} 2003), or a
high value of $E_c$ (Zhang and Huang 2004).

Another case providing insight is that of a smooth shifted power-law
$G(E)=A(E+E_*)^{-\delta-1}$, which has no edges though the
corresponding $F(E)$ has a smooth peak at $E=E_*/\delta$. In this
case the fractional error in $G$ due to applying (22) ignoring the
recombination term can be expressed as

\begin{equation}
\frac{\Delta G(E)}{G(E)} = (\delta+1)\frac{D}{E+E_*}\Sigma_{Z_{eff}}Z_{eff}^4A_{Z_{eff}}\left[\frac{1}{1-V_{Z_{eff}}/(E+E_*)}\right]^{\delta+2},
\end{equation}
where each term in the $Z_{eff}$ sum is zero for
$E<V_{Z_{eff}}=Z_{eff}^2\chi$.

In the case of recombination onto Fe XXV alone (hot plasma),
this gives for $\delta =5$,

\begin{equation}
\frac{\Delta G}{G}\approx \frac{10 ~ keV}{E+E_*}\left[1-7 ~
 keV/(E+E_*)\right]^{-7},
\end{equation}
 which is shown in Figure 6 for $E_*=5,10,20$ keV.
 Evidently errors due to neglect of recombination can be large
 at low $E$. The reason is that the $Z_{eff}$ recombination contribution
 to the bremsstrahlung solution for $G(E)$ at $E$ comes from the slope of
 $G$, and not just $G$ itself and at $E-V_{Z_{eff}}$ not at $E$.
Figure 6 is similar to Figure A.2 because $F_2/F_1=G_2/G_1= 1+\Delta
G/G_1$.

 This error has very serious consequences for past
analyses of HXR flare spectra, at least in cases where a significant
hot dense coronal loop is involved. For example, the f-b emission
spectrum is most important at lower energies (5-30 keV or so),
depending on the plasma temperature $T$ and low energy electron cut-off or roll-over $E_c, E_*$ and is steeper than the free-free. This
will offset some of the spectral flattening caused around such
energies by photospheric albedo (Alexander and Brown 2003, Kontar {\it et
al.} 2005) resulting in underestimation of the albedo contribution and
hence of the downward beaming of the fast electrons. This fact would
weaken the finding of Kontar and Brown (2006) that the electrons are
near isotropic, in contradiction of the usual thick target
description, but for the fact that the flares they used had rather
hard spectra and substantial footpoint emission - conditions where
the f-b correction should be rather small. Nevertheless it
illustrates that care is needed to ensure f-b emission is properly
considered.

Finally, recognising the presence of the f-b contribution, one can
in fact convert integral Equation (20) into a
differential/functional equation for $F(E)$ by differentiation,
namely

\begin{equation}
G(E) -
D\Sigma_{Z_{eff}\ge{E/\chi}^{1/2}}A_{Z_{eff}}Z_{eff}^4G'(E-Z_{eff}^2\chi)
= -H'(E),
\end{equation}
which is a wholly new class of functional equation in need of
exploration.

\section{Discussion and Conclusions
}

It is clear from our findings that ignoring non-thermal f-b
contribution as negligible, as has been done in the past, is
erroneous. Even if we ignore coronal enhancement of element abundances, and use photospheric abundances, f-b contribution can be very significant. In certain flaring regions, especially in dense-hot
coronal sources or occulted loop-top events, fast electron
recombination can be of vital importance in analysing data properly
and in inferring electron spectra and energy budgets. It can have a
major influence on inferred electron spectra both as an inverse
problem and also in forward fitting parameters, including the
important potential to find and evaluate low-energy electron
cut-offs, which are vital to flare energy budgets. While
incorporating f-b into spectral fitting procedures will make it
considerably more complicated, an advantage is that the f-b, unlike
the f-f, contribution retains its $J(\epsilon)$ signatures of any
sharp features in $F(E)$.

A major consequence of the low energy f-b contribution is that, to
fit an actual photon spectrum, less electrons are needed, than in
f-f only modelling, at the low $E$ end, which is where most of the
power in $F(E)$ lies. For example, if we consider the case
$\delta=5,E_c=10$ keV and ionisation up to Fe XXV, then we see
from Figures 1 and 2 that inclusion of f-b increases $j$ by a factor
of 2-10 in the 15-20 keV range for $\delta=$ 3-5. Thus, to get a
prescribed $j$ in that range we need only $10-50\%$ as many
electrons as inferred from f-f emission only.

We also note that the importance of non-thermal f-b emission is
greatest when non-thermal electrons are present at low $E$ and with
large $\delta$ such as in microflares with 'hard' XRs in the few to
ten KeV range (Krucker {\it et al.} 2002). Such low energy electrons
have short collisional mfps and so are more likely to emit mainly in
hot coronal regions, if accelerated there. Microflares are therefore
important cases for inclusion of f-b.

Before we conduct any precise fitting of $F(E)$, involving the f-b
contribution, to real data (e.g. from RHESSI) and include it in
software packages it will be important to include, for both f-b and
f-f, more accurate cross-sections with Gaunt factors etc. and
ionisation fractions as functions of plasma temperature. By doing this, it
will be possible to show, for certain events, how vital
recombination is and to improve our understanding of electron
spectra and their roles in flares. However, our Kramers results
already bring out the fact that recombination should not be ignored
in the future, and that it may be invaluable in some cases as a
diagnostic of the presence or otherwise of electron spectral
features.

\section*{Acknowledgements}
This work was supported by a PPARC Rolling Grant and UC Berkeley
Visitor funds (JCB) and by a Dorothy Hodgkin's Scholarship (PCVM).
Helpful discussions with A. Caspi, H.S. Hudson, A.G. Emslie and J.
Kasparova are much appreciated, as are the helpful suggestions of the referee (S. Krucker).

\vskip 0.5cm

\appendix
\vspace{-0.5cm}
\section{Efficiency and smooth $F(E)$}

\begin{figure*}
\centering
\includegraphics {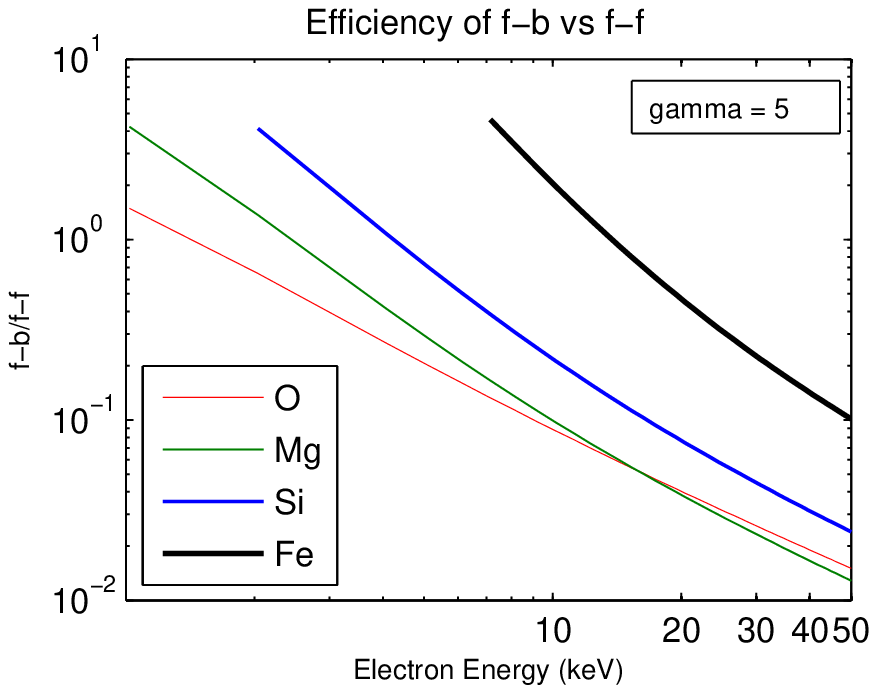}
\caption{The f-b electron efficiency compared to f-f for the 4
elements discussed in Appendix A. It is evident from the graph
that, if present, highly ionised Fe is the most efficient
source of f-b HXRs in terms of the $F(E)$ needed followed by Si, O
and Mg.}
\end{figure*}

\begin{figure*}
\centering
\includegraphics {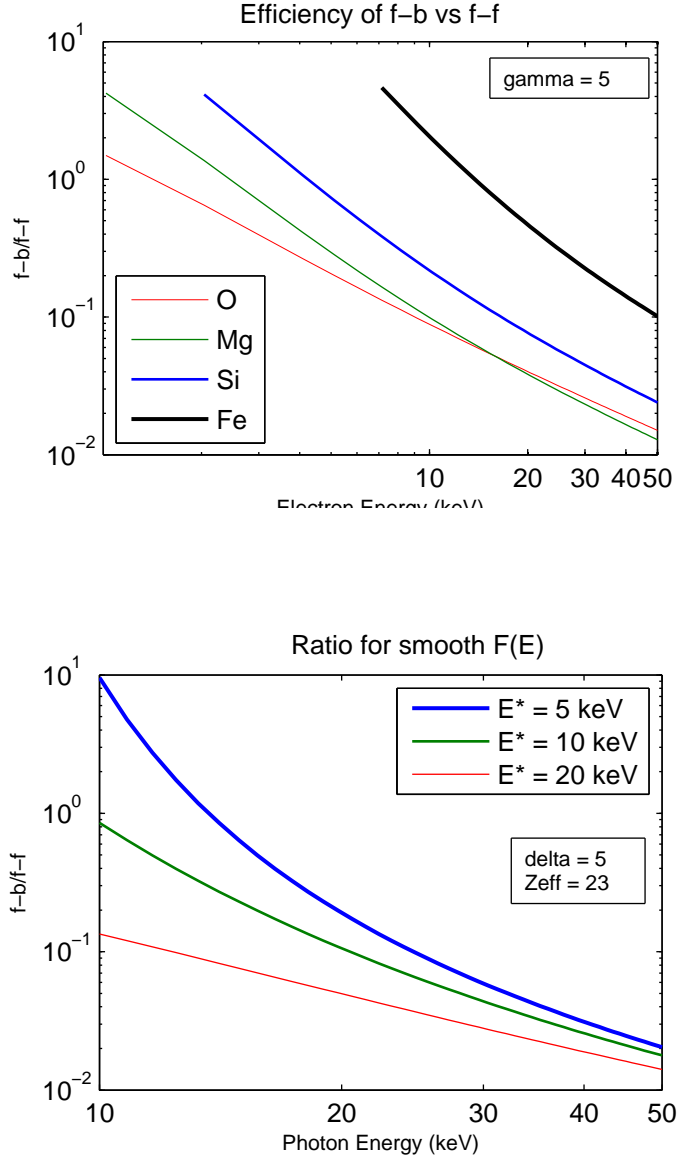}
\caption{The $\Psi_{smooth}$ as discussed in Appendix A and Equation (A.5). It is the ratio of $j_R$ to $j_B$ for the
smooth $F(E) \propto E(E+E_*)^{-\delta-1}$ for $E_*=5, 10, 20$ keV
respectively.}
\end{figure*}

\subsection{Comparison of the efficiency of f-b versus f-f HXR yield}
In Section 3.1 and 3.2, we predicted the $j_B, j_R$ from a power-law $F(E)$ and found
that the $j_R$ contribution could sometimes be more important than
$j_B$. It is of interest therefore, to consider the following
question. If one observes a power-law $j(\epsilon)\propto
\epsilon^{-\gamma}$ above some $\epsilon \ge V_Z$, what electron
flux $F_R(E)$ would be needed to generate it in a plasma of solar abundances {\it purely} by
non-thermal electron recombination on ion $Z_{eff}$ as compared with
the $F_B(E)$ required to do so {\it purely} by f-f bremsstrahlung? If we write, from Equation (15), the
latter as $F_B(E)=CE^{-\gamma+1}$ then the former has to be, by
Equation (16),
\begin{equation}
F_R(E)=C(E+V_{Z_{eff}})^{-\gamma}/DZ_{eff}^4A_{Z_{eff}},
\end{equation}
where
\begin{equation}
D=\frac{2\pi \chi}{{\sqrt 3}\zeta_B}\approx 0.04 keV
\end{equation}
and the ratio measuring recombination efficiency relative to bremsstrahlung is
\begin{equation}
\frac{F_B(E)}{F_R(E)}=\gamma
Z_{eff}^4A_{Z_{eff}}\frac{D}{E}[1+V_{Z_{eff}}/E]^{\gamma},
\end{equation}
which we show in Figure A.1 for $\gamma=5$ in terms of each of the dominant f-b contributions
from fully ionised O, Mg, Si and Fe respectively while the f-f is for all elements. Evidently
non-thermal recombination could be dominant over bremsstrahlung up
to many 10s of keV as the most efficient HXR source if the electrons are
emitted entirely in a plasma hot enough ($T\approx 20$MK) for
elements up to Fe 24+ to be ionised and is significant even at lower temperatures.

In terms of the total required electron fluxes $F_{R1},F_{B1}$ above
energy $E_1$, the ratio is

\begin{eqnarray}
\nonumber \frac{F_{B1}}{F_{R1}}& =\frac{\gamma-1}{\gamma-2}
Z_{eff}^4
A_{Z_{eff}}\frac{D}{E_1}[1+V_{Z_{eff}}/E_1]^{\gamma-1}&\\
\approx & 0.02
Z_{eff}^4A_{Z_{eff}}\frac{10~keV}{E_1}[1+V_{Z_{eff}}/E_1]^{\gamma-1},&
\end{eqnarray}
which is about 10 for Fe, 0.25 for Si and 0.1 for Mg and O at
$E_1=10$ keV.

At higher electron energies ($E \ge \approx 17$ keV), O becomes more efficient than Mg, as can be seen
in Figure A.1, because of the combined effects of the $A_ZZ^4$ factor and the term containing $V_Z$.

\subsection{Ratio of $j_R$ to $j_B$ for an example of a smooth $F(E)$ with no cut-off}
All of the above results are for $F(E)$ with a sharp cut off $E_c$.
To illustrate how the appearance of $j(\epsilon)$ is modified by
inclusion of f-b as well as f-f for a smooth $F(E)$, a simple case to
evaluate is $F(E)\propto E(E+E_*)^{-\delta-1}$, which behaves as
$E^{-\delta}$ at $E\gg E_*$ but has a smooth roll-over at
$E_*/\delta$. It is simple to show that the resulting
$j_B(\epsilon)\propto (E+E_*)^{-\delta}/\delta$ for f-f alone and
that the ratio of f-b to f-f in this case is, for ion $Z_{eff}$
alone,

\begin{equation}
\nonumber \Psi_{smooth} = \frac{D\zeta_{Z_{eff}}}{\epsilon+E_*}\left[1-\frac{Z_{eff}^2}{\epsilon+E_*}\right]^{-\delta-1},
\end{equation}
which is shown in Figure A.2 for $\delta=5$, $Z_{eff}=23.77$ and
$E_*=5, 10, 20$ keV. We see again that $\Psi_{smooth}$ is largest for large $\delta$ and for small $E_*$.

\vskip 0.5cm

\vspace{-0.5cm}

\section{Whole Flare Thin Target, Thermal, and Thick Target Expressions
for f-f and f-b HXR Emission Spectra}

\begin{figure*}
\centering
\includegraphics{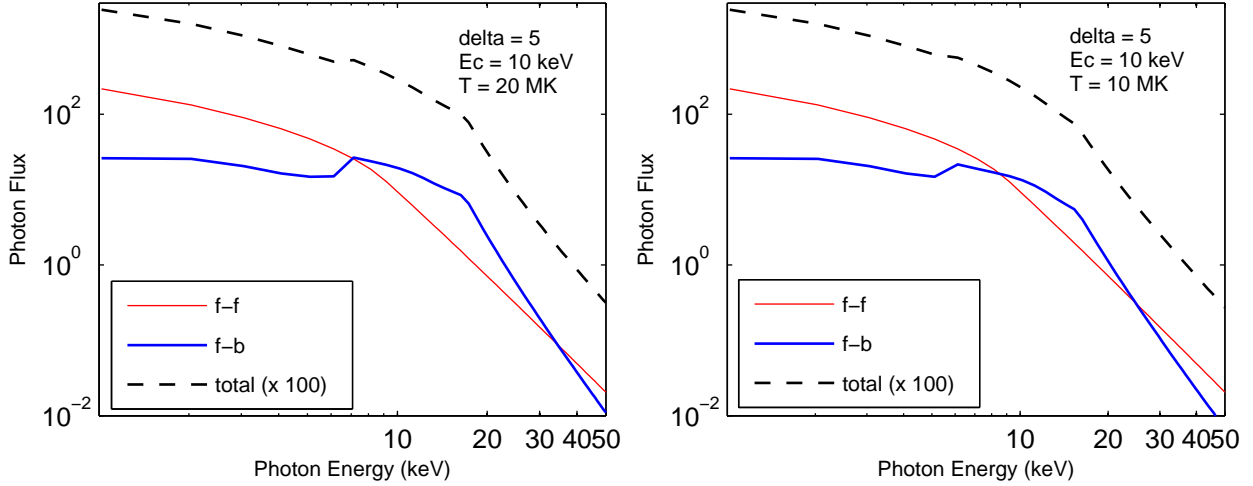}
\caption{Non-thermal f-f and f-b spectra for the thick
target case (Equations (B.11) and (B.12)) shown for 2 different temperatures: 20 MK that is pertinent to events such as the 2002 April 14 event and 10 MK, which is more in the range of 'microflare' temperatures. It is
interesting to note the three distinct energy regimes for the f-b
spectrum, namely: $\epsilon < V_{Fe}; V_{Fe} \le \epsilon \le V_{Fe} + E_c; \epsilon > V_{Fe} + E_c$. Clearly f-b is very important in the 10-50 keV range,
precisely where albedo issues are also important.}
\end{figure*}

Here we extend the above results on local emissivities $j(\epsilon)$
to estimate total spectral emission rate $J(\epsilon)$
(photons sec$^{-1}$ per unit $\epsilon$) from extended flare volumes as required for real flare data.

\subsection{Thin Target Coronal Loop}
A thin target is one in which $F(E)$ is not significantly modified
by energy losses or gains over the volume. For a loop of half length
$L$, transverse area $A$, volume $2AL$ and density $n_p$, the total emission rate
spectra contributions $J_{thin}(\epsilon)$ are for a power law $F(E)$ with a low energy cut-off, by Equation (15),

\bigskip
\begin{eqnarray}
\nonumber J_{B_{thin}}(\epsilon)  = &
\frac{\delta-1}{\delta}\frac{8\alpha\zeta_B}{3}\frac{m_ec^2r_e^2}{\epsilon}
\frac{2n_pALF_c}{E_c}\times &\\
\nonumber & ~~\left[\frac{\epsilon}{E_c}\right]^{-\delta}; &
\epsilon \ge E_c \\ & ~~1; & \epsilon < E_c
\end{eqnarray}
and by Equation (16),
\begin{eqnarray}
\nonumber J_{R_{thin}}(\epsilon) =& (\delta-1)\frac{32\pi
\zeta_{RZ_{eff}}}{3^{1/2}\alpha}\frac{r_e^2
\chi}{\epsilon}\frac{2n_pALF_c}{E_c^2} \times \Sigma_{Z_{eff}}& \\
\nonumber & \times ~~\left[\frac{\epsilon-Z_{eff}^2\chi}{E_c}
\right]^{-\delta-1}; & \epsilon \ge E_c+Z_{eff}^2\chi \\
\nonumber & \times ~~0; & \epsilon  < E_c+Z_{eff}^2\chi, \\&&
\end{eqnarray}
where the summation is over all $Z_{eff} \le (\epsilon-E_c]^{1/2}$. These spectral shapes $J(\epsilon)$ are of course just the same as
the thin target $j$ forms, scaled by the plasma volume.

\subsection{Hot Coronal Loop Thermal Emission (in the Kramers approximation)}

Both f-f and f-b emissions are included in the standard analyses
(e.g. Mewe {\it et al.} 1987, Dere {\it et al.} 1996) of isothermal
hot plasma contributions to flare spectra, using full cross sections
and ionisation balance expressions. It is therefore surprising that
f-b is omitted from calculations of non-thermal emission, especially
at low $\epsilon$, where electrons of comparable energy are present
in both thermal and non-thermal populations. In applying our study
of the non-thermal f-b to real data we wish to include thermal
emission as it is important at energies under about 20 keV and so
dilutes the visibility of non-thermal contributions. In order to
treat the thermal and non-thermal $j$ consistently and allow
meaningful comparisons we use the expressions for the thermal $j$
relevant to the Kramers cross sections just as in the non-thermal
case - but see remarks previously and below concerning Gaunt factors
and absolute accuracy of our results.

For an isothermal plasma the local Maxwellian electron flux spectrum
is

\begin{equation}
F_{therm}(E)=\left[\frac{8}{\pi m_e}\right]^{1/2}
\frac{E}{(kT)^{3/2}}n_p\exp(-E/kT),
\end{equation}
which, by Equation (4), gives for the thermal bremsstrahlung emission
from a uniform loop

\begin{equation}
J_{Btherm}(\epsilon) = \frac{16\alpha r_e^2}{3}\zeta_B m_ec^2 \times
\left[\frac{8}{\pi m_e}\right]^{1/2}\frac{2n_p^2AL
e^{-\epsilon/kT}}{\epsilon(kT)^{1/2}}
\end{equation}
and for the recombination

\begin{eqnarray}
\label{} \nonumber J_{Rtherm}(\epsilon) &= \sqrt{\frac {2\pi}{27
m_e}}\frac {64r_e^2\chi^2}{\alpha}  \frac {2n_p^2AL}{\epsilon
(kT)^{3/2}} \Sigma_{Z_{eff}} \zeta_{RZ_{eff}} & \times \\ &
\exp\left(\frac{Z_{eff}^2\chi - \epsilon}{kT}\right).&\\ \nonumber
\end{eqnarray}

These results can be compared with those of Culhane (1969) and
Culhane and Acton (1970) who were among the first to explicitly
address the X-Ray spectrum from hot coronal plasmas. Using the
Kramers cross sections is essentially equivalent to setting to unity
all Gaunt factors in their expressions. When we do so, the
$\epsilon, T$ dependences of our $ J_{Rtherm}, J_{Btherm}$ are
identical to theirs - e.g. $ J_{Rtherm}/ J_{Btherm}$ is independent
of $\epsilon$, the only difference being that our $J_{Rtherm}$ is
much larger (in absolute value) than theirs, mainly because they
used the very much lower value of $A_Z$ for Fe believed at that
time. Examination of the $\epsilon, T$ dependences of Culhane's
Gaunt factors shows that they affect quite significantly both the
f-f and the f-b  spectra from a Maxwellian $F(E)$ and we should
expect the same to be true for non-thermal $F(E)$ like power-laws.
Thus, any accurate absolute comparison of predictions with data will
require incorporation of appropriate $g,G$. However, these do not
affect the absolute orders of magnitude of $J_{Rtherm}, J_{Btherm}$
nor the dependencies on $n_p,V,F_c$ etc., nor the locations of
edges. So, for the present purpose of demonstrating the importance
of f-b, the Kramers expressions will suffice.

\subsection{Thick target (dense loop or footpoint) f-f and f-b emission
spectra}

In the thick target case, $j$ evolves in space along with the energy
losses of the electrons. To find $j$ locally one uses the continuity
equation (Brown 1972) and then integrates over volume to get $J$.
However, to get the whole volume $J$, it is actually simpler (Brown
1971)  to start with the electron injection rate spectrum ${\cal
F}_o(E_o)$ electrons/sec per unit injection energy $E_o$ and use the
expression

\begin{equation}
J_{thick}(\epsilon) = \int_{E_o} {\cal F}_o(E_o) \eta (\epsilon,
E_o)dE_o,
\end{equation}
where $\eta (\epsilon, E_o)$ is the total number of photons per unit
$\epsilon$ emitted by an electron of energy $E_o$ as it decays in
energy. For purely collisional losses $dE/dN=-K/E$ with $K=2\pi e^4
\Lambda$, $e$ being the electronic charge and $\Lambda$ the Coulomb
Logarithm. Then

\begin{equation}
\eta(\epsilon,E_o) =\frac{1}{K}\int_E E\frac{dQ}{d\epsilon} dE
\end{equation}
for the relevant radiation cross section $dQ/d\epsilon$. Note that
this assumes H to be uniformly and fully ionised along the electron
path. For partially ionised H the energy loss constant $K$ is
reduced but this situation is not relevant to our hot source
situations.

For our Kramers $dQ/d\epsilon$ f-f and f-b expressions (3), (7) and (9), the resulting expressions, in the case where
$A_{Z_{eff}}$ are uniform along the path, Equation (B.7) gives

\begin{eqnarray}
\nonumber \eta_B(\epsilon,E_o) =& \frac{8\alpha\zeta_B}{3}\frac{r_e^2m_ec^2}{K} & \times \\
\nonumber & \left[\frac{E_o}{\epsilon}-1\right]; & \epsilon \le E_o \\
& 0; & \epsilon > E_o
\end{eqnarray}
and

\begin{eqnarray}
\nonumber \eta_{RZ}(\epsilon, E_o) & = \frac{32\pi
A_{Z_{eff}}Z_{eff}^4}{3^{3/2}\alpha}\frac{r_e^2\chi^2}{K\epsilon} &\times\\
\nonumber & 1; & E_o\ge\epsilon+Z_{eff}^2 \chi \\
& 0; & E_o <\epsilon+Z_{eff}^2.
\end{eqnarray}

 For a power-law injection rate spectrum
 of spectral index $\delta_o$, viz

\begin{equation}
{\cal F}_o(E_o) = (\delta_o-1)\frac{{\cal
F}_{oc}}{E_{oc}}\left[\frac{E_o}{E_{oc}}\right]^{-\delta_o};   E_o \ge E_{oc},
\end{equation}
where  ${\cal F}_{oc}$ is the total rate above low energy cut-off
$E_{oc}$, the expressions for the non-thermal emission spectra are
then by Equation (B.6)

\begin{eqnarray}
\label{} \nonumber J_{Bthick}(\epsilon) =& \frac{8\alpha
r_e^2}{3}\frac{\zeta_B m_ec^2{\cal F}_{oc}}{(\delta_o - 1)(\delta_o
-2)K} & \times \\
\nonumber & \left(\frac
{\epsilon}{E_c}\right)^{-\delta_o +1}; & \epsilon \ge E_c \\
\nonumber  & \left[(\delta_o-1)\frac{E_c}{\epsilon}-(\delta_o-2)\right]; &\epsilon < E_c\\
\end{eqnarray}
and, for ion $Z_{eff}$,

\begin{eqnarray}
\label{}
 \nonumber J_{RZ_{eff}thick}(\epsilon) = & \frac{32\pi
r_e^2
m_ec^2}{3^{3/2}\alpha}\zeta_{RZ_{eff}}\frac{\chi^2}{K\epsilon}\frac{{\cal
F}_{oc}}{E_{oc}}  \times  &\\
\nonumber &
\left[\frac{\epsilon-Z_{eff}^2\chi}{E_{oc}}\right]^{-\delta_o + 1}; &
\epsilon \ge E_{oc}+Z_{eff}^2\chi \\
 \nonumber & \left[\frac{E_{oc}-Z_{eff}^2\chi}{E_{oc}}\right]^{-\delta_o + 1}; & Z_{eff}^2\chi<\epsilon <
 E_{oc}+Z_{eff}^2\chi\\
& 0; & \epsilon < Z_{eff}^2\chi.
\end{eqnarray}

For the case of a cold thick target footpoint the total $\zeta_R$
can be almost as small as 1 if only hydrogen and some low $\zeta_R$
elements are ionised and even zero if $T<8000$ K or so (there being
almost no charged ions present). In these sources the f-b
contribution is negligible or at most a very small correction. For a
collisonally thick hot loop $\zeta_R$ is, however, very much higher.

The main distinction of these hot thick target spectra compared to
hot thin targets is that the decay of all electrons to zero energy
means that the signature of the cut off $E_{oc}$ in the injection
spectrum appears not as a discontinuity in $J(\epsilon)$ but only in
its gradient $J'(\epsilon)$. This gradient break is very noticeable
in Figure B.1 at energy $\epsilon = E_c+V_{Fe}$. So, even in the thick
target case, spectral diagnosis of any $E_{oc}$ present is possible.
The recombination edges themselves appear at the relevant ionisation
energies $\epsilon = V_{eff}$, these being from thick target electrons
decelerated to zero $E$. These non-thermal recombination spectral
edges are then down in the energy regime below 10 keV which is
complicated by Fe lines etc., making the interpretation of
${\cal F}_o$ there, and of the lines, more difficult.

\end{document}